\documentclass[prb,preprint,showpacs,floatfix]{revtex4}

\usepackage{graphicx} 
\begin{document}

\title{Quasiparticle electronic structure of charged oxygen vacancies in TiO2 }
\author{Ali Kazempour}
\author{Javad Hashemifar}
\author{Hadi Akbarzadeh}
\affiliation{Department of Physics, Isfahan University of Technology,Isfahan 84156-83111 , Iran}

\begin{abstract}
We studied the  oxygen vacancies($V_{O}$)  in rutile TiO2 by using $G_0W_0$ approximation on top of GGA+$U$  as a method of choice to improve the gap. Since there is  no extensive agreement regarding the characteristic of electron localization for $ TiO_2$, we examine combined $G_0W_0$@GGA+$U$ scheme in which both  are conceptually one step toward better enumeration of the non locality of exchange-correlation potential . Our $G_0W_0$@GGA+$U$ results realize and confirm the weak nature of electron correlation in rutile $TiO_2$ and shows that the $U$-dependence of the energy gap in perfect bulk is slightly stronger than in defected sample. In addition, we studied the $U$- dependency of $V_O$ defect states and found that different charged vacancies  shows different $U$-dependence . While the application of $G_0W_0$  correction would improve the  quasiparticle  gap and formation energies, however the $V_O$ states , in contrast to experiment, remains entangled with the conduction band. Finally, we used PBE0 and HSE06 hybrid functionals and found that these exchange-correlation functional particularly HSE06 that reproduced the real gap and provide desire description of the screening , properly disentangle the neutral and singly ionized $V_O$ from the conduction band . According to hybrid functional calculations, all vacancies  are stabilized  and  $V_O$ with  2+ charge state is the most stable vacancy  in the whole Fermi-level range inside the gap and hence $V_O$ acts as a shallow donor.

\end{abstract}

\pacs{
}

\maketitle
$TiO_2$ has attracted many interests with application as photo catalysis, water splitting, solar cell and  sensors\cite{1,2}. It has been demonstrated that $TiO_2$ can be reduced easily and reach high degree of nonstoichiometry which leads to high n-type conductivity \cite{3}. With thermal excitation, oxygen vacancy ($V_O$) may loose its electrons and becomes charged. Some thermogravimetric measurements  reported that the electrical conductivity
demonstrate various  scaling as a function of oxygen partial pressure and temperature . Each scaling behavior correspond to the dominance of solely one vacancy  charge state \cite{3}.
On the other hand , there are experimental findings that shows rather dominance of Ti interstitial and independent picture of conductivity in terms of oxygen partial pressure \cite{4}.
On the theoretical side,several investigations have been performed from embedded cluster to ab-initio total energy methods on the oxygen-deficient rutile $TiO_2$ sample.\cite{chen,chao,ramam}. Cho et al.\cite{chao} using local-density approximation (LDA) observed the entangled $V_O$ state inside conduction band.  Park et al.\cite{park}  in an alternative ab-initio study using Hubbard correction   for both $Ti-3d$ and $O-2p$ were only able to locate $V_O^+$ below conduction band minimum about 1 eV . But it doesn't give the neutral vacancy $V_O^0 $ state in the gap.
Moreover both above mentioned theoretical studies suffer from the band gap underestimation. In spite of  wide research on rutile including the role of $V_O$ in n-type electrical conductivity, its localized nature is still under debate. Therefore it seems necessary to reinvestigate the features of $V_O$ in $TiO_2$.
Although Density Functional Theory has been regarded as a valuable tool for microscopic understanding of defect mechanisms, but the inadequate description of self-energy and underestimation of energy gap in popular local density (LDA) and generalized gradient approximation (GGA) leads to a widespread deficiencies.
These limitations lead to artificial characterization of defect  states and wrong position of  thermodynamic transition levels. It is believed that overcoming gap problem leads to reliable determination of defect states motivated using  methods such as  DFT+$U$, Hybrid functionals and Self-interaction correction to partly improve the self-interaction. Qualitatively, Many Body Perturbation Theory (MBPT) in the $GW$ approach has shown successful treatment of quasi-particles band structure  in semiconductors and is considered as a promising method to study defects \cite{5}. However, application of $GW$ on systems with partially filled d or f orbital due to wrong-predicted LDA band ordering  yield unreliable electronic spectra  and the dependence of spectra on adjustable parameters or starting point remain involved  \cite{6}. Alternatively,  LDA or GGA starting point might be substituted with one giving closer features to respective QP band energies.
As a benchmark, we choose GGA+$U$ single-particle solution as starting point in  $GW$ scheme to evaluate problematic positioning of $V_O$ in reduced $TiO_{2-x}$. Moreover,this choice may provide insightful  information about localized nature of the $V_O$ states. Our results show smoothly $U$-dependence of vacancy affinity  all are grater than CBM affinity  and therefore $G_0W_0$@GGA+$U$  fails to stabilize $V_O^0$ and $V_O^+$ inside the gap.
Additionally, in the realm of post-DFT methods, we employ Hybrid functionals PBE0 and HSE in which they partially reduce self-interaction errors and improve gap in the description of vacancy states. The obtained results show that $V_O$ acts as a shallow donor with $V_O^{2+}$ being the most stable charge in the whole variation of Fermi level.
\section{Computational details}
The calculations are performed using GGA+$U$ based on PBE formalism for generalized gradient approximation \cite{pbe} as starting point for $G_0W_0$ and PBE0 and HSE06 exchange-correlation functional with a plane wave basis set and norm-conserving pseudopotentials.
Atomic orbitals 3s, 3p, 4s, and 3d  are included into the Ti valence
subspace, while 3s and 3p orbitals are considered as valence for O.
 Spin-polarized calculation are considered for systems with an odd number of electrons. The cutoff energy 70 Ryd was chosen based on convergence of the cohesive energy of rutile $TiO2$ and for Brillouin zone integration we use a mesh of $2\times 2\times 2$. To remove the interaction of neighboring supercell we performed 72, 96 and 108 atoms supercell calculations in which the  difference between 72 and 96 atoms supercell defect formation energy is less than 0.02 eV and 0.04 eV for neutral and charged vacancies and therefore choose 72 atoms supercell. All defected supercells are fully relaxed  until the forces are below 1 mRyd/bohr. In GGA+$U$, the value of the Hubbard term $U$ , applied to the Ti 3d states, is varied between 0 and 4 eV. This range was chosen based on the estimate of the value of $U$ obtained using a constrained GGA calculation, as implemented in the QUANTUM ESPRESSO code \cite{7},and also by calculating $U$ for an isolated Ti atom, divided by the theoretical optical dielectric constant $ \epsilon _{\infty }$= 6.7. Both approaches give similar $U$ values of about 1.2 eV.
 The $GW$ approach is applied non-self-consistently ($G_0W_0$ ), exploiting the first order expansion of the self-energy \cite{GW1,GW2}. The dynamic behavior of the dielectric matrix is determined by the plasmon-pole approximation as implemented in the SAX code \cite{8}. The defect formation energies are calculated according to
\begin{equation}
E^f[X_{q}]=E_{\rm tot}[X_{q}]-E_{\rm tot}[{\rm bulk}]+\mu_O+q[E_{\rm F}+E_{\rm VBM}+\Delta V]
\end{equation}
where $E_{\rm tot}[X_{q}]$ and $E_{\rm tot}[{\rm bulk}]$ are defected  and perfect supercell total energy, respectively. The reference for the oxygen chemical potential is chosen to be 1/2$EO_2$ , while for Ti the chemical potential is referenced to the bulk Ti
. The electronic chemical potential(Fermi energy) is referenced to the valence band maximum (VBM) corrected by the alignment of the electrostatic potential in perfect  and defect supercell far enough from defect states. Moreover ,in Charged cases, the Makov-Payne correction \cite{9} have not been included due to  large dielectric constant entering in Madelung term.
\section{ $G_0W_0$@GGA+$U$ Methodology}
It was argued that GGA+$U$ can be viewed as an approximation to $GW$ for localized $d$ and $f$ states \cite{10}. However , due to $U$ correction on localized states,the hybridization of localized states with others has not been included i.e. the itinerant states remain at GGA level that in some cases isn't true  . Moreover , in GGA+$U$, the double counting (in the LDA+$U$ formalism , double counting removes a part of electron-electron interaction that was already included in LDA Hamiltonian)  term is not well defined. On the other hand screening is described statically while in fact it behaves dynamically and has strong energy dependence for particularly localized electrons  . Furthermore it was shown that for open-shell and shallow d systems the application of $G_0W_0$ on GGA doesn't yield  fine QP band structure attributed to wrong GGA-derived band ordering \cite{11}. One solution here is to replace the GGA with GGA+$U$ method which its eigenvalues and eigenfunctions are relatively closer to QP electronic structure \cite{12,13}.
Despite GGA+$U$, $G_0W_0$ calculation based on GGA+$U$ has not shown $U$-dependence explicitly while the resulted QP state would show implicit $U$-dependence \cite{13}.
In the formalism of GGA+$U$ based $G_0W_0$ \cite{13}, the QP equation is expressed as:
\begin{equation}
\varepsilon_i \approx \epsilon^{ks}_i +\rm Z_i<\phi^{ks}_i|\Sigma(\epsilon_i^{ks})-V_{xc}- \rm V_{db}|\phi^{ks}_i>
\end{equation}

which $V_{db}$ accounts for double counting term. The GGA+$U$ single particle equation is:
\begin{equation}
\epsilon_i^{ks}=<\phi_i^{ks}|-\frac{1}{2}\nabla^2 +V_{GGA}+V_{db}|\phi_i^{ks}>=\overline\epsilon_i^{ks}+<\phi_i^{ks}|V_{db}|\phi_i^{ks}>
\end{equation}
where $\overline\epsilon_i$ can be considered as GGA energies obtained with GGA+$U$ wave function. Eventually, with linear expansion of self-interaction operator $\Sigma(\epsilon_i^{ks})$ around $\overline\epsilon_i$ the QP equation can be written as:
\begin{equation}
\varepsilon_i=\overline\epsilon_i^{ks}+Z'_i <\phi_i^{ks}|\Sigma(\overline\epsilon_i^{ks})-V_{xc}|\phi_i^{ks}>
\end{equation}
which clearly  demonstrate advantage of having no double counting term . Regarding the first term,  $U$ correction affects not only the energies of the levels, but also the hybridization between the orbitals, if allowed by symmetry. The most pronounced modification are expected to appear near the Fermi level, where a significant change in localization of electronic states can occur. The states deeper in energy are already quite localized and atomic-like, so that a non-zero $U$ would mostly result to  a shift in their energies. In this case, the localized $d$ or $f$  occupied (unoccupied) states are pushed toward lower (higher) energies  and get more localized. But the situation for half-filled states are quite demanding and severely depends on the nontrivial hybridization of included orbital as a function of $U$ and it's character.  The second term would results in screening alteration and get tendency to increase energy gap with $U$ addition. In general both terms of above equation have contributed into $U$-dependence of the system in $GW$@GGA+$U$ scheme.

\section{results and discussion}
\subsection{A.$GW$@GGA+$U$}
In previous section , we reviewed the application of $GW$@GGA+$U$ to determine the affinity and  ionization energy i.e. band gap of the perfect system in which the CBM is fully empty. In the case of defected sample which defect-induced state lies inside the gap or in resonance with band edges, the situation  slightly changed so that the highest occupied states might be the defect state (in the case the defect states falls inside the gap) or CBM (in which the defect states is in resonance  or above CB). It was shown that for $TiO_2$ all charged $V_O$ states becomes merged in CB in GGA or even GGA+$U$ derived band structure. However, the Motivation here for the choice of GGA+$U$ is that the application of $U$ would decreases the outward relaxation of vacancy neighbors and enhance their hybridization leads to better description of the defected sample band structure closer to its real quasiparticle spectra. 

\begin{figure}[ht]
\includegraphics{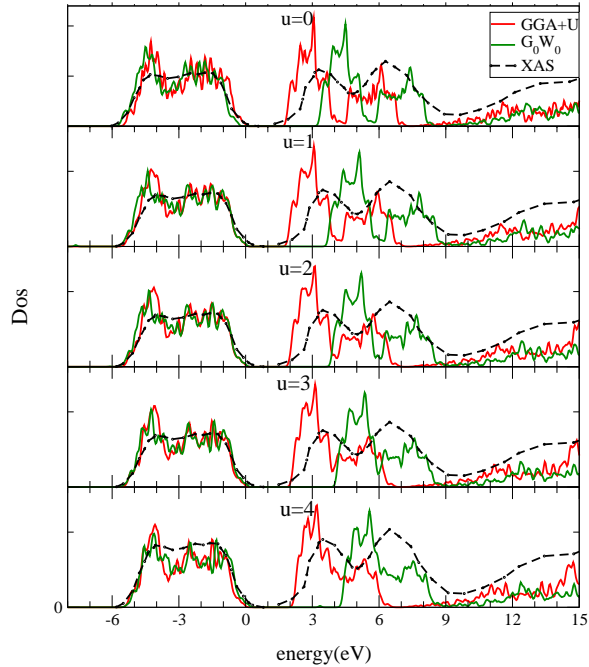}
\caption{\label{gw-gga-bulk} The Dos of $TiO_2$ from GGA+$U$, $G_0W_0$@GGA+$U$ and XAS are compared within the range of $U=0,4$ .}
\end{figure}

Prior to studying defected material, it is advantageous  to inspect $G_0W_0$@GGA+$U$ electronic structure of the perfect bulk. Fig.\ref{gw-gga-bulk} shows comparison between GGA+$U$, $G_0W_0$@GGA+$U$ and  XAS experimental density of states of bulk $TiO_2$ \cite{14}. Experimental XAS and GGA+$U$  was aligned at upper valence band edge with $G_0W_0$ spectrum.  The sharp peak of empty $Ti-3d$  in XAS spectra is situated between its analogous in GGA+$U$ and $G_0W_0$ . While the addition of $U$ would enlarge the GGA+$U$ gap  and the $3d$ peak of $Ti$ becomes closer to XAS counterpart, $G_0W_0$ gives overestimated gap with $3d$ peak gradually goes away from XAS correspondent.

\begin{figure}[ht]
\includegraphics{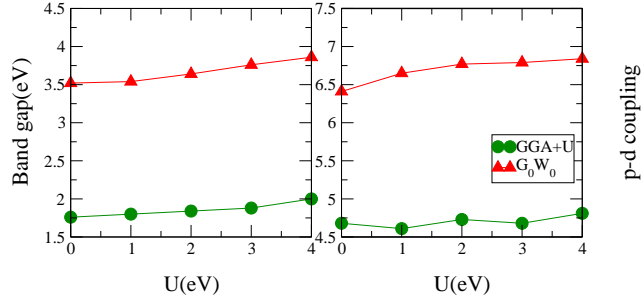}
\caption{\label{perfect-gap} The values of band gap (left) and the p-d peak distance are shown as a function of $U$ from GGA+$U$ and $G_0W_0$@GGA+$U$.}
\end{figure}

 In both GGA+$U$ and $G_0W_0$ the  most significant correction occurs for CB that are made of $3d$-Ti via reduction of  the conduction band width toward higher energies . As it is shown in  Fig.\ref{perfect-gap}  the $U$-sensitivity is more pronounced in $G_0W_0$ than GGA+$U$. The reason could be expressed in response to the question that how much the effect of $U$  change single particle wave-functions and screening in dressed potential that consequently reflect the degree of $p-d $ coupling . It realize that the hybridization of localized orbital with itinerant states  that is neglected in GGA+$U$, play an important role when combined with $G_0W_0$ . In other word, the differences in sensitivity of energy gap with  $U$ values could be traced back to the augmentation of localized and  itinerant hybridization as a function of $U$. From the inspection of p-d  coupling it is evident that smooth behavior of $U$-sensitivity   of the gap confirm the weakly correlated nature of the $TiO_2$  that already  was obtained about 1.2 eV in previous section. Another useful criteria derives from $U$ dependence of electronic structure is  the dimensionless  $U/W$ ratio that shows the degree of the localization   in $TiO_2$  typically varies between 0.5-2 for the selected range of $U=1,2,3,4$(W is the $3d-Ti$ bandwidth) . By comparison with the typical $U/W$ values obtained for highly correlated material such as $NiO$ which  exceed from 4 \cite{13} , it illustrate the weak localization of  d-electrons in $TiO_2$.

\begin{figure}[ht]
\includegraphics[scale=0.35]{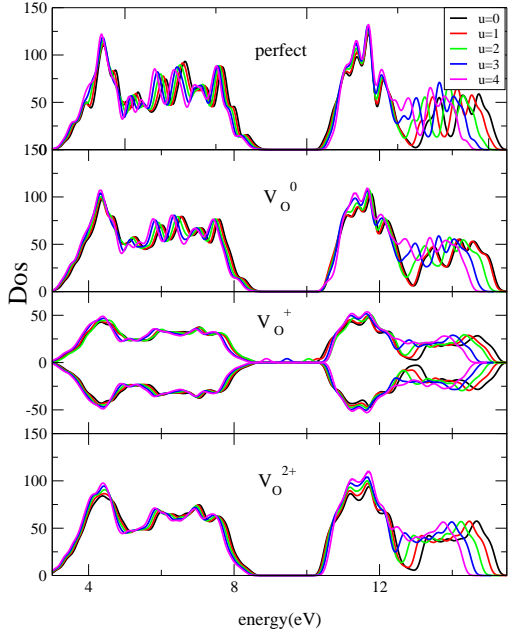}
\caption{\label{dos-gga-u} The effect of $U$ correction on the electronic structure of
perfect bulk and different charged oxygen vacancy is compared. For $V_O^+$ with $U$ addition the defect induced state appear in below CBM.}
\end{figure}

In the case of defect involved $TiO_2$, with the addition of $U$,  GGA+$U$ resulted  $V_O^0$ and $V_O^{2+}$ energy states  remain within the conduction band whereas for $U=2,3,4$ , $V_O^+$ state appear below CBM with the values 0.73, 1.12 and 1.73 eV ,respectively.  Fig.\ref{dos-gga-u} shows the expected downward shift of fully occupied CBM in $V_O^0$ and  half-filled defect states in $V_O^+$ and upward shift of  unoccupied CBM  in $V_O^+$ and   $V_O^{2+}$ though the gap problem exist. Having earned the $V_O$ states of GGA+$U$, we are able to evaluate the influence of hubbard correction $U$ compared to $GW$@GGA+$U$ one. 

\begin{figure}[ht]
\includegraphics[scale=0.35]{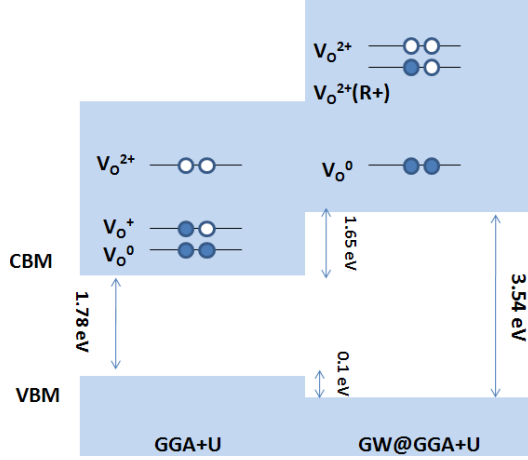}
\caption{\label{state-gw-gga+u} Single particle GGA+$U$ (left) and quasi-particle energies (right) for symmetric defect state of various $V_O$ are shown. The self-energy correction would shift up all defect states specially singly and doubly ionized vacancies.}
\end{figure}

\begin{figure}[ht]
\includegraphics[scale=.8]{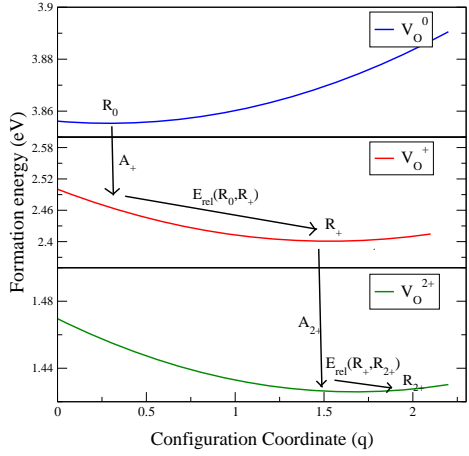}
\caption{\label{cc} The simple calculated configuration coordinate diagram with GGA method. Each charged vacancy was computed at three configuration ($R_0$, $R_+$ and$ R_{2+}$). The diagram show closer values of $A_{2+}$ and $I_+$ for singly and doubly ionized vacancies.}
\end{figure}

The $G_0W_0$@GGA+$U$ corrected quasi-particle states  are shown for all three charged states of $V_O$ relative to initial GGA+$U$ in Fig.\ref{state-gw-gga+u}. Since the most $GW$ codes does not allow a direct treatment of systems with unpaired electrons, the $V_O^+$ defect affinity is estimated using the relation  A(+/2+) = −I (2 + /+)  where A and I are affinity and ionization of given defect state for local charge addition. For this reason, we have performed a $G_0W_0$ calculation with doubly ionized vacancy at the geometry of relaxed $V_O^+$ . Calculated configuration coordinate diagram is shown in Fig.\ref{cc} also clearly indicate almost closer configuration of $V_O^{2+}$ and $V_O^+$ and proximate same values of I(+/2+) at $R_+$ and $R_{2+}$   .
 The affinity of the whole of the system was measured relative to upper valence band edge VBM while  the affinity of defect states was measured with respect to the lower conduction band CBM. For each $V_O$, $G_0W_0$ corrected states shifts up within the conduction band.

\begin{figure}[ht]
\includegraphics[scale=0.5]{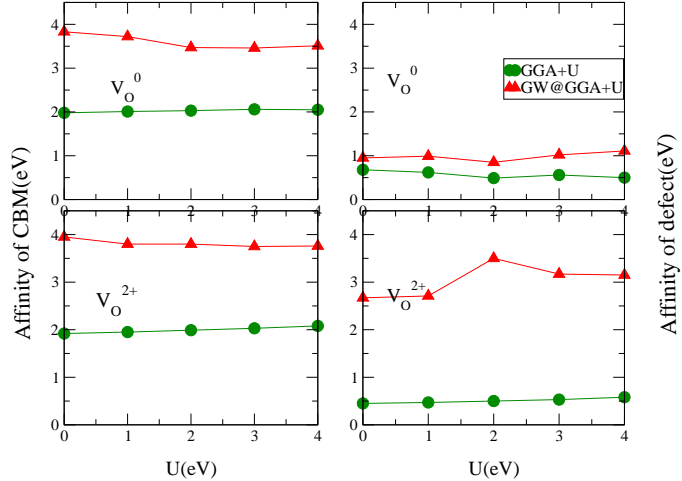}
\caption{\label{affinity} (left) and (right) represent the comparison of affinity of the whole system and the defect state affinity, for $V_O^0$ (top) and $V_O^{2+}$ (bottom) , respectively.}
\end{figure}

As a result,it is observed that empty $V_O^{2+}$ and half-filled $V_O^+$ is more sensitive to the self-interaction correction than filled $V_O^0$ and the shift of their quasiparticle energies is more remarkable than $V_O^0$.
Fig. \ref{affinity} shows the affinity of CBM and vacancy state of $V_O^0$ and $V_O^{2+}$ species. In $G_0W_0$@GGA+$U$, the affinity of CBM .i.e band gap for $V_O^0$ and $V_O^{2+}$ decrease with increasing $U$ within 0.35 eV and 0.20 eV  while this values amount to 0.10 eV and 0.15 eV in GGA+$U$, respectively. For the vacancy affinity the resulted values for $V_O^0$ and $V_O^{2+}$ varies within 0.16 eV and 0.48 eV for $G_0W_0$  and 0.18 eV and 0.13 eV for GGA+$U$, respectively.
Compared to GGA+$U$ , it is apparent that  the U dependence of $G_0W_0$ is slightly more for the gap and vacancy affinity . Moreover, the affinity of the oxygen vacancy in both GGA+$U$ and $G_0W_0$ shows similar linear ascending trend whereas  the trend for band gap show inconsistent behavior. Inspection of the quasi-particle correction, shows that the upward shift is more prominent for  $V_O^{2+}$ than $V_O^0$ . This indication illustrate the different characters for  vacancy  level  in which the $G_0W_0$ application  effectively change those states concerned  vacancy center with no electrons(with no cost of electrostatic energy). At $U$=2 both vacancies involved concrete change in quasi-particle states. This , however is an artifact of GGA+$U$ that shows new band ordering and the vacancy level crossover some states due to change in its hybridization..

Using $GW$ correction scheme was proposed by Rinke {\it{et al}}\cite{15} based on separation of formation energy into local electron addition $A$ and lattice relaxation part $\Delta$, we obtained the corrected formation energy and thermodynamic transition levels. At the first step, the vertical electron affinity between two charged states at fixed geometry is calculated by $G_0W_0$ formalism . In a second step, the lattice relaxation  are measured after charge addition between initial and final geometries with GGA+$U$.(Fig .\ref{cc})

Starting from reference formation energy of $V_O^{2+}$, the corrected formation energies of $V_O^+$ and $V_O^0$ are then given by
\begin{equation}
E_{f}(+,\epsilon_{f})=\Delta(+,R+,R2+)+A(2+,R2+)+E_{f}(2+,\epsilon_{f} =0)
\end{equation}

and
\begin{equation}
E_{f}(0,\epsilon_{f})=\Delta(+,R0,R+)+A(+,R0)+E_{f}(+,\epsilon_{f} =0)
\end{equation}
 Using calculated affinities for all defects involved we calculated formation energies as  a function of $U$ summarized in table.\ref{cor-formation}.
\begin{table*}
\caption{\label{cor-formation}All corrected formation energies by $GW$@GGA+$U$ are listed in units of eV. The starting formation energy of V$^{2+}_{o}$ has been taken from GGA+$U$ values. }
\begin{tabular}{cccccccccc}
\hline \hline
   &$ \Delta_{2} $ & $ \Delta_{1}$  &  A(+/0) & A(2+/+)
   & $E_{f}(2+)$ & \multicolumn{2}{c}{$E_{f}(+)$} & \multicolumn{2}{c}{$E_{f}(0)$} \\
\hline
  & & & & & &GGA+$U$ & G0W0 & GGA+$U$ & G0W0  \\
  \hline
   U=0  & 0.05 &0.01 & 0.75   &2.66 & 0.66 & 2.16 & 3.37 & 3.98 & 4.14\\
   U=1  & 0.25 &0.04 & 0.79  & 2.71 & 0.85 & 2.33 &3.81 & 4.25 &4.64  \\
   U=2  & 0.36 &0.08 & 0.67  & 2.43 & 1.00 & 2.41 &3.79 & 4.53 &4.54   \\
   U=3  & 0.47 &0.10 & 0.82   & 2.87 & 1.22 & 2.42 &4.56 & 4.88 &5.48  \\
   U=4  & 0.55 &0.15 & 0.90  & 2.96& 1.42 & 2.37 &4.93 & 5.02 & 5.98  \\
\hline
\end{tabular}
\end{table*}

\begin{figure}[floatfix]
\includegraphics[scale=1.2]{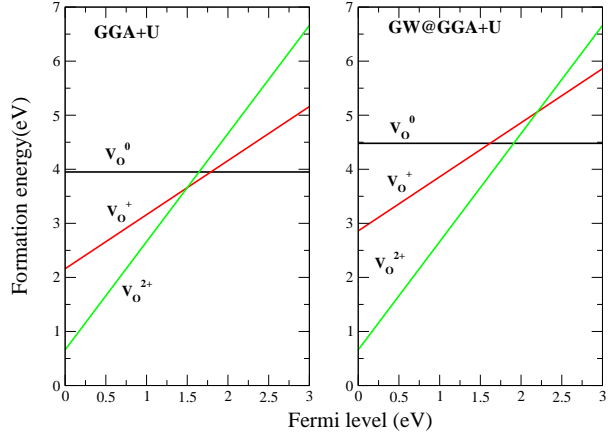}
\caption{\label{transition} The comparison between GGA+$U$ and correction scheme using $G_0W_0$@GGA+$U$ is shown. $G_0W_0$ correction method results in metastable $V_O^+$ with variation of Fermi energy.}
\end{figure}

The resulted formation energy for $V_O^+$ varies within 1.2-2.5 eV and   within the rang of 0.15-0.95 eV for $V_O^0$ as a function of $U$. The increase of $V_O^+$ formation energy relative to neutral ones implies that its concentration is now much lower than the GGA+U counterpart as is shown in Fig.\ref{transition} . Also affinity of $V_O^+$ changes inasmuch as its stability as a variation of Fermi level (chemical potential) in the gap is vanished relative to GGA+$U$ . With  $U$ addition, this trend will continue up along with the $V_O^{2+}$ becomes the most covered stable charge in the energy gap and consequently the charged transition level $\epsilon(2+/0)$ would shift up though remain inside the gap. Furthermore, the charged transition level $\epsilon(2+/0)$  change from 1.70 and 1.89 eV at $U=1$ up to 1.82 and 2.28 eV at $U=4$ for GGA+$U$ and $G_0W_0$ , Respectively. Hence, although $G_0W_0$@GGA+$U$ doesn't stabilize $V_O^0$ and $V_O^+$ in the gap,it gives $V_O^{2+}$ the most dominant stable vacancy with growing $U$ and eliminate the stability of singly ionized $V_O$ for all fermi-level position within the gap.

\subsection{B.Hybrid functionals, HSE06 and PBE0}
We now turn to hybrid functionals that hopefully overcome the band gap problem. While a number of hybrid functional have been introduced \cite{16}, we already focus on
$PBE0$ and $HSE06$ which typical gap has qualitative agreement  with experiment. While recently published study with HSE06 functional gives comprehensive description of oxygen vacancies \cite{hse-jannoti}, the comparison of HSE06 and PBE0 results gives efficient insight how does the short and long range of exchange interaction affect the defect levels. In principle ,for both PBE0 and HSE06 ,the exchange energy part is taken as

\begin{equation}
E_x^{hyb}=aE_x^{HF}+(1-a)E_x^{GGA}
\end{equation}
with a=0.25 and the correlation part remains at GGA level except that in the latter the short-range part of the HF exchange potential is kept \cite{17}.

\begin{figure}[floatfix]
\includegraphics[scale=1.1]{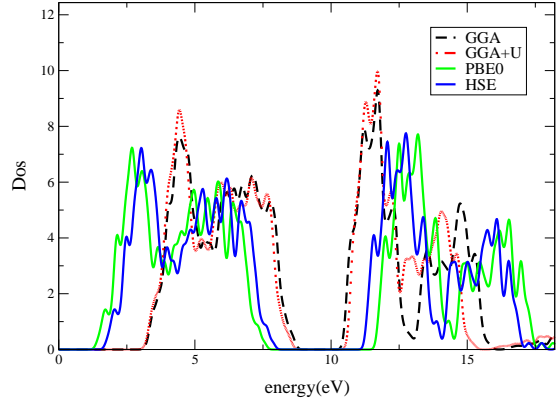}
\caption{\label{dos-compare} The DOS of hybrid functionals HSE06 and PBE0 compared to GGA and GGA+$U$ are plotted. A portion of HF exchange addition in hybrid functionals give raise to gap opening. }
\end{figure}

 The reason behind choosing these functionals lies in the fact that they provide improved gap in many materials \cite{18}. The calculations performed in this section was done with same parameter and accuracy as  in previous section. Furthermore, our obtained total energy include singularity correction of the G=0 exchange potential .  For the perfect $TiO_2$ bulk the energy gap are 4.15 eV and 3.10 eV for PBE0 and HSE06 ,respectively shows good agreement of HSE06 gap with experimental value 3.05 eV. The comparison of different functional gap are shown in Fig.\ref{dos-compare}.
from inspection of above density of states it seems that the addition of a portion of HF exchange would considerably increase the gap relative to GGA. The comparison of  PBE0 and HSE06 also reflect the fact that the energy gap is correctly reproduce  by the short range nature of exact exchange. The VBM state lowering
and CBM state lifting amount to 0.96 eV and 1.19 eV for PBE0 and 0.63 eV and 0.77 eV for HSE06. This trends is attributed to the HF admixed portion which reduce the self-energy of VBM and also raise the CBM.
Using above methods we calculated the energetic, electronic and structural properties of each vacancy type. Since the vacancy induced state are made from the three Ti atoms surrounding the vacancy , their relaxation magnitude is  determinant to the positioning of vacancy states in the band structure. Table. \ref{bond}represent the distances of vacancy neighbors for GGA, PBE0 and HSE06.

\begin{table}
\caption{\label{bond} The magnitude of vacancy neighbors outward relaxation in $\AA$ for PBE0 and HSE06 are listed. The relative comparison shows lower relaxation of fully and half occupied vacancy levels(due to bounded electron to defect center) in hybrid functionals than GGA. }
\begin{tabular}{ccccccc}
\hline\hline
Ti-O &\multicolumn{2}{c}{$V_O^0$}& \multicolumn{2}{c}{$V_O^+$}& \multicolumn{2}{c}{$V_O^{2+}$} \\
\hline\hline
GGA & 1.128& 1.151 & 1.127& 1.153  & 1.126& 1.157 \\
PBE0 &1.014& 1.120 & 1.042& 1.076  & 1.130& 1.163 \\
HSE06 &1.000& 1.121 & 1.009& 1.124 &1.149& 1.175\\
\hline\hline
\end{tabular}
\end{table}

\begin{figure}[floatfix]
\includegraphics{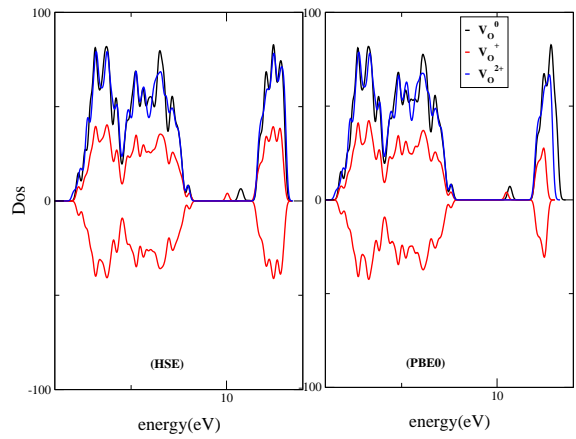}
\caption{\label{pbe0-hse-dos} Both HSE06 and PBE0 are successful in giving the neutral and singly ionized vacancies below CBM . HSE06 band gap is coincide with experimental gap and also the defect  state is in closer agreement with experimental values and Ref\cite{hse-jannoti}.  }
\end{figure}

It is obvious that in PBE0 and HSE06 the outward relaxation for $V_O^0$ and $V_O^+$ that have bounded electrons to vacancy center is relatively small compared to GGA while bigger outward relaxation is observed for $V_O^{2+}$ than GGA analogous due to having no bounded electron.
Having earned the amount of vacancy relaxation, we are able to interpret the obtained states in the gap as shown in Fig.\ref{pbe0-hse-dos}.

\begin{figure}[floatfix]
\includegraphics[scale=1.1]{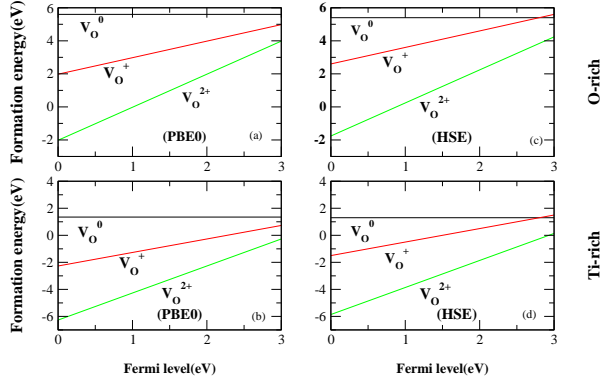}
\caption{\label{trans} The diagram of formation energy with Fermi energy variation is depicted for PBE0 and HSE06 hybrid functional for O-rich and Ti-rich conditions. Both functionals give the $V_O^{2+}$ the stable vacancy with Fermi energy change and transition level $\epsilon(2+/0)$ is approximately within $\sim$ 1 eV above conduction band minimum. }
\end{figure}

Upon introducing oxygen vacancy , its neighbors relax in order to reinforce their bonding with the rest of the lattice and results in less overlap between
three Ti atoms and shift the vacancy state up. On the other hand if the vacancy center has bounded electrons it resist against outward relaxation since it gives unstable energetic picture. Hence balance of these  two competitive factors determine the  position of vacancy states. From the table \ref{bond}, it evidences that HSE06 and PBE0 show less  outward relaxation than GGA for $V_O^0$ and $V_O^+$ while the situation is reversed for $V_O^{2+}$  which give raise to the $V_O^0$ and $V_O^+$  levels 1.12 and 1.25 eV for PBE0 and 0.80 and 1.40 eV for HSE06 below CBM .  Comparing   $V_O$ states for neutral and singly ionized charged vacancy in PBE0 and HSE06 indicate that energy shift of $V_O^0$ state is more than $V_O^+$ due to the presence of short part of exact exchange. Finally we address the extrapolated formation energy at O-rich and Ti-rich limit as shown in Fig. \ref{trans}. The formation energy of $V_O^{2+}$ is fairly low compared to $V_O^+$ and $V_O^0$ in terms of Fermi energy variation in the gap . The reason lies in the fact that these hybrid functionals lower the VBM level considerably  lead to lowest formation energies(with no electron bounded to vacancy) and therefore the  charge transition level lies above the gap . Our results for formation energy and $\epsilon(2+/0)$ is closed to HSE results of Ref.\cite{hse-jannoti} . This implies that both $V_O^0$ and $V_O^+$  have always higher formation energy than $V_O^{2+}$ make it only possible stable charged state of oxygen vacancy. This behavior is exactly same in both O-rich and Ti-rich regime except that the formation energies in Ti-rich are much lower than O-rich. Accordingly, this observation is mainly consistent with experimental measurement suggest conductivity reduction with increase of oxygen partial pressure \cite{kofsted}.

\section{summary}
In conclusion, we studied the problem of oxygen vacancies in viewpoint of gap correction method. With the application combined $GW$ correction and GGA+$U$ starting point referred as $GW$@GGA+$U$, we found that the quasi particle energies shows different $U$-dependence traced back to their character and occupancy.. Additionally, we found that application of $G_0W_0$ depends on the starting point .  It also fails to predict the vacancy state inside the gap contrary to the experimental findings. Alternatively , we employed hybrid functional PBE0 and HSE06 in which they are able to truly predict the neutral and singly ionized vacancies in the energy gap . However they gives raise to stable energetic picture of doubly ionized vacancy as a function of Fermi energy variation within the gap and transition level $\epsilon(2+/0)$ about 0.6 eV and 0.8 eV for PBE0 and HSE06 above CBM reflect shallow n-type conductivity features of $V_O$.
\section{acknowledgment}
This work was supported partially by the Vice Chancellor for Research Affairs of Isfahan University of Technology and ICTP Affiliated Center. A.Kazempour gratefully acknowledge fruitful discussions with M. Scheffler, S. Levchenko and P.Rinke. A. Kazempour appreciate FHI Institute der Max Plank Gesellschaft for financial support and computation resources during his visit.
\bibliography{TiO2.bib}

\end{document}